\documentclass[english]{extarticle}

\usepackage[T1]{fontenc}
\usepackage[latin9]{inputenc}
\usepackage{color}
\usepackage{babel}
\usepackage{amsmath}
\usepackage{amsthm}
\usepackage{amssymb}
\usepackage{setspace}
\usepackage[unicode=true,pdfusetitle,
 bookmarks=true,bookmarksnumbered=false,bookmarksopen=false,
 breaklinks=true,pdfborder={0 0 1},backref=false,colorlinks=true]
 {hyperref}

\makeatletter
\numberwithin{equation}{section}
\numberwithin{figure}{section}
\newcommand{\lyxaddress}[1]{
	\par {\raggedright #1
	\vspace{1.4em}
	\noindent\par}
}

\@ifundefined{date}{}{\date{}}
\makeatother

\begin{document}
\title{\textbf{On the quaternion transformation and field equations in curved
space-time}}
\author{\textbf{B. C. Chanyal}}
\maketitle

\lyxaddress{\begin{center}
\textit{Department of Physics, G. B. Pant University of Agriculture
\& Technology}\\
\textit{Pantnagar-263145 (Uttarakhand) India. Email: bcchanyal@gmail.com}
\par\end{center}}
\begin{abstract}
In this paper, we use four-dimensional quaternionic algebra to describing
space-time geometry in curvature form. The transformation relations
of a quaternionic variable are established with the help of basis
transformations of quaternion algebra. We deduced the quaternionic
covariant derivative that explains how the quaternion components vary
with scalar and vector fields. The quaternionic metric tensor and
geodesic equation are also discussed to describing the quaternionic
line element in curved space-time. Moreover, we discussed an expression
for the Riemannian Christoffel curvature tensor in terms of the quaternionic
metric tensor. We have deduced the quaternionic Einstein\textquoteright s
field-like equation which shows an equivalence between quaternionic
matter and geometry.

\textbf{Keywords:} quaternion, curvature space-time, Riemannian Christoffel
curvature tensor, Einstein field equation

\textbf{PACS:} 02.40.Hw, 02.40.Ky, 4.20.-q

\textbf{Mathematics Subject Classification 2020: }12Hxx, 35Q76, 83C10
\end{abstract}

\section{Introduction}

The theory of the physical world has two fundamental components: matter
and its interactions. The fundamental forces or interactions can be
described as a field in which particles interact with one another.
These fundamental interactions broke down into four fundamental forces
of nature \cite{key-1}: the gravitational force due to the attraction
between massive bodies, the electromagnetic force due to the force
between electrically charged particles, the weak force responsible
for the radioactive decay of atoms, and the strong force that binding
fundamental particles together. Gravitational force is the weakest
force experiences as attractive. Because it is a very long-range force,
the dominant force on the macroscopic scale is the reason for the
formation, shape, and orbit of the astronomical bodies. This force
is responsible for an object in space to move in a circular orbit.
In Newtonian physics, the concept of absolute space and time is important
for the dynamics of an object but they are not relative to each other.
Further, Albert Einstein developed the theory of relativity, therefore,
the traditional Newtonian concept of absolute space and time has been
replaced by the concept of space-time. Time is considered as the fourth
dimension in the four-spaces theory of special and general relativity.
The theory which explains the gravitation is named \textquoteleft General
Theory of Relativity\textquoteright{} (GTR) \cite{key-2}. In GTR,
gravity does not act as a force but significance the space-time curvature
where the source of the curved space-time is energy and matter. The
Einstein field equations also elaborate that the mass and energy are
responsible for the curvature of space-time, and the geometry of space-time
is responsible for the motion of matter. Since, the description of
space-time geometry should be non-Euclidean geometry, such that the
space-time path becomes curved in presence of the gravitational field.
In this case, the path of motion of particles in such a field or non-Euclidean
space-time shows curvilinear instead of straight. Objects in nature
have basic properties that are independent of the manual selection
of any specific coordinate system. The basic assumption of physics
is that the laws of nature are stated by equations, and should be
valid for all frames of reference.

The theory of relativity required the \textquoteleft covariant form\textquoteright{}
that is independent of the coordinate system. The covariant form remains
the same in all co-ordinate systems whether the reference frame will
be changed or not. The laws of physics must be expressed by an equation
that remains covariant for all coordinate systems. The tensor has
such properties that show covariant form in all coordinate systems.
If all the equations, related to the general theory of relativity,
are written in the tensor form, then it will transform easily from
one reference frame to another. Therefore, in mathematics, the tensor
calculus is the best tool to explain the theory of curvature as ordinary
calculus. Besides, for a flat space-time, the Pythagoras theorem is
valid perfectly but for a curved space-time, this theorem no longer
holds and uses a \textquoteleft metric\textquoteright{} to measure
the distance. Moreover, in GTR, the metric tensor behaves as the gravitational
potential of Newtonian gravitation. The Riemannian Christoffel's curvature
tensor captures the notion of parallel transport. It gives the idea
about the connections of the parallel transported vector on a curve
space-time. Further, Riemannian curvature is required for all the
changes in the tangent vectors when we transport them around a curved
space. The Riemann-Christoffel tensor is the only tensor that can
be constructed from the second derivative of the metric tensor \cite{key-3}.
The contraction of the Riemann tensor gives Ricci curvature and scalar
curvature. Ricci curvature explains the concept in which mass converges
and diverges in time corresponding to the part of the curvature of
space-time. Scalar curvature gives the single real number which represents
the quantity of Riemannian manifold volume of geodesic ball differs
from the normal ball in Euclidean space. The equation of motion of
free particles is recognized by geodesic equation followed the analogous
to Newton\textquoteright s equation of motion which clarifies the
equation for the acceleration of particles. The source of the curvature
of space-time is generalized by the energy and momentum of the field
expressed as an energy-momentum tensor.

The study of algebraic structure in curvature space-time has an important
role to describe the natural world. In mathematical algebraic structure,
there are four types of division algebra \cite{key-4}, i.e., the
real algebra, complex algebra, quaternion algebra (Hamilton algebra)\cite{key-5},
and octonion algebra (Cayley algebra)\cite{key-6}\cite{key-7}\cite{key-8}.
Real numbers are the numbers, which are used normally without any
imaginary number. The numbers written as the mixture of a real number
and the imaginary number are renowned by complex numbers. Quaternions
are the extension of a complex number with non-commutative property,
used to label the rotations in three-dimension. Cayley algebra is
the extension of the quaternions with non-commutative and non-associative
as well. There are many applications of quaternionic algebra in various
branches of theoretical and computational physics \cite{key-9}\cite{key-10}\cite{key-11}\cite{key-12}\cite{key-13}.
In electrodynamics, Maxwell\textquoteright s equations in the presence
of magnetic monopoles \cite{key-14} and the classical wave equations
of motion \cite{key-15} have been constructed in terms of quaternionic
algebra. The quaternionic form of quantum electrodynamics has been
discussed \cite{key-16} \cite{key-17}. Chanyal \cite{key-18}\cite{key-19}
proposed the quaternionic covariant theory of four-dimensional particle
dyons in the form of relativistic quantum mechanics and also focused
on the quantized Dirac-Maxwell equations for dyons. Recently, in magneto-hydrodynamics,
the quaternionic dual fields equations for dyonic cold plasma have
been analyzed \cite{key-20}. Further, it has been developed Dirac-Maxwell,
Bernoulli, and Navier Stokes like equations for dyonic fluid-plasma
in the generalized quaternionic field \cite{key-21}. Currently, a
new approach to Dirac's relativistic field equation for rotating free
particle has been investigated in quaternionic from \cite{key-22}.
Beyond the quaternionic algebra, many authors \cite{key-23}\cite{key-24}\cite{key-25}\cite{key-26}\cite{key-27}\cite{key-28}\cite{key-29}\cite{key-30}\cite{key-31}\cite{key-32}\cite{key-33}\cite{key-34}\cite{key-35}
have studied the role of higher dimensional hypercomplex division
algebras in various fields of modern physics. In GTR, Edmonds \cite{key-36}
discussed the quaternion wave equation in curved space-time by using
the curvilinear coordinate system in relativistic quantum theory.
The quaternionic form of curvature quantum theory fills the gap between
quantum and gravity theory. In the same way, Weng\textbf{ }\cite{key-37}
studied the electromagnetic and gravitational field equations in complex
curved space with the help of quaternions and octonions. Keeping in
view the Riemannian space and its connection with quaternions, we
discussed the Einstein field-like equation in quaternionic curvilinear
form. Starting with quaternionic basis transformation from one frame
to another, we define the transformation of scalar and vector field
components of a quaternion curvature variable. From the quaternionic
covariant derivative, we expressed the Christoffel symbol, quaternionic
metric tensor, and quaternionic geodesic equation. We also formulated
the quaternionic Riemannian tensor that keeps track of how much scalar
and vector components of quaternion changed if we propagate parallel
along with a small parallelogram. Interestingly, if quaternionic Riemannian
Christoffel curvature is zero then the quaternionic curved space-time
is converted into flat space-time. Further, the quaternionic Ricci
tensor is an important contraction of the quaternionic Riemannian
Christoffel tensor which explains the changes in four space-time when
an object parallels transport along a geodesic. We have discussed
the quaternionic form of the Einstein field-like-equation in a compact
form which shows that the quaternionic value of matter-energy is equivalent
to the quaternionic geometry.

\section{Preliminaries }

A tensor is the generalized form of a vector, defined by simply the
arrangement of numbers, or functions that transform according to certain
rules under a change of coordinates \cite{key-38} \cite{key-39}.
According to Einstein summation convention, if any same index appears
twice in a term then that index stands for the sum of all terms at
the complete range of values. For example, we can write the two forms
of a vector $\mathbf{u}$ as
\begin{align}
\mathbf{u}\,=\,\,\sum_{i}u^{i}e_{i}\,\,, & \,\,\,\,\,\,\,\,\,\mathbf{u}\,=\,\,\sum_{j}u_{j}e^{j}\,\,\,\,\,\,\,\,\,\,\,\,\,\,\,\,\,\,\,\,\,\,\,\,\forall\,\,(i,\,j=\,\,1,2,3)\,,\label{eq:1}
\end{align}
where $i$ and $j$ are the repeated indices or summation indices
while $u^{i}\,\text{and\,}u_{j}$ are the vector components in contravariant
and covariant form, respectively. The scalar product of two vectors
can be represented in terms of indices with their components, i.e.,
\begin{align}
\mathbf{a}\cdot\mathbf{b}\,\,=\,\, & \left(\sum_{i}a_{i}e_{i}\right)\cdot\left(\sum_{j}b_{j}e_{j}\right)\,=\,\,\sum_{ij}a_{i}b_{j}\left(e_{i}\cdot e_{j}\right)\,=\,\,\sum_{ij}a_{i}b_{j}\delta_{ij}\,=\,\,\sum_{i}a_{i}b_{i}\,,\label{eq:2}
\end{align}
where $\delta_{ij}$ is the Kronecker delta symbol defined as $\delta_{ij}=\,\,1,\,\text{for}\,(i=\,j)$~and~$\delta_{ij}=\,0,\,\text{for}\,(i\neq\,j)$.
Also, the vector product becomes
\begin{align}
\left[\mathbf{a}\times\mathbf{b}\right]_{i}\,\,=\,\, & \sum_{jk}\epsilon_{ijk}a_{j}b_{k}\,.\label{eq:3}
\end{align}
Here $\epsilon_{ijk}$ is the Levi-Civita symbol with three indices
having value $\epsilon_{ijk}=\,+1$ for cyclic permutation, $\epsilon_{ijk}=\,-1$
for non-cyclic permutation, and $\epsilon_{ijk}=\,0$ for any two
repeated indices. The tensors recognized by their order viz. vectors
are first-order tensors, dyadics are second-order tensors, triadics
are third-order tensors and tetradics are fourth-order tensors. By
adding any two tensors of the same rank it gives the tensor of the
same order, i.e.,
\begin{align}
\mathbf{A}_{st}^{r}\,\,=\,\, & \mathbf{B}_{st}^{r}+\mathbf{C}_{st}^{r}\,,\label{eq:4}
\end{align}
where $\mathbf{A}_{st}^{r},\,\mathbf{B}_{st}^{r}$ and $\mathbf{C}_{st}^{r}$
are the tensors of same rank, $i.e.$ three-rank tensor. The product
of two tensors is given by
\begin{align}
\mathrm{B}_{s}^{r}\,\mathbf{C}_{uv}^{t}\,\,=\,\, & \mathcal{A}_{suv}^{rt}\,.\label{eq:5}
\end{align}
In the above representation, $\mathrm{B}_{s}^{r}$ is tensor of rank
two having indices $r$ and $s$; $\mathbf{C}_{uv}^{t}$ is a tensor
of rank three having indices $t,u\text{ and }v,$ while $\mathcal{A}_{suv}^{rt}$
is a tensor of rank five. If a magnitude or scalar quantity $\phi$
transform from one reference frame to another, it remains invariant,
such that \cite{key-40}
\begin{align}
\phi^{'}\,\,=\,\, & \phi\,.\label{eq:6}
\end{align}
This invariant quantity may also be known as contravariant tensor
of rank zero or covariant tensor of rank zero. On the other hand,
a vector quantity can be transformed as
\begin{align}
v^{'r}\,:\longmapsto\,v^{s},\,\,\,\,\Longrightarrow\,\,\,\,v^{'r}\,\,=\,\, & \frac{\partial x^{'r}}{\partial x^{s}}v^{s}\,.\label{eq:7}
\end{align}
Equation (\ref{eq:7}) signifies that the components of vector $v^{s}$
transformed to components of vector $v^{'r}$ when the coordinate
$x^{s}$ transformed to $x^{'r}.$ We can also define the contravariant
tensor components of rank two as
\begin{align}
\mathrm{A}^{'ru}\,:\longmapsto\,\mathrm{A}^{st},\,\,\,\,\Longrightarrow\,\,\,\,\mathrm{A}^{'ru}\,\,=\,\, & \frac{\partial x^{'r}}{\partial x^{s}}\frac{\partial x^{'u}}{\partial x^{t}}\mathrm{A}^{st}\,.\label{eq:8}
\end{align}
Similarly, we can transform the tensors for higher rank.

\section{The quaternionic algebra}

In mathematics, a complex number over the real algebra $\mathbb{R}$
having the imaginary number $i$, is defined by the algebraic extension
of a normal real number. Any complex number $Z\in\mathbb{C}$ can
be expressed in the form of its basis $\left(1,i\right)$ as
\begin{align}
Z\,\,= & \,\,\xi_{1}+i\xi_{2},\,\,\,\,\,\,\,\,\,\,\,\,\,\,\,\,\,\,\,\,\,\forall\,\left(\xi_{1},\,\xi_{2}\right)\in\,\mathbb{R}^{2}\,\,\text{and }\,Z\in\,\,\mathbb{C}\,.\label{eq:9}
\end{align}
If $Re\left(Z\right)=0$ then the complex number is called purely
imaginary. In the same way, the quaternion is a number system that
extends the complex number in a form of algebra having some properties
over addition and multiplication.\textbf{ }W. R. Hamilton \cite{key-5},\textbf{
}extended the complex number in terms of four-dimensional norm-division
algebra (quaternion) over real algebra $\mathbb{R}.$ The quaternions
has four unit elements $\left(e_{0},e_{1},e_{2},e_{3}\right)$ called
basis elements, in which $e_{0}$ is the scalar unit and $e_{1},e_{2},e_{3}$
are the imaginary units. A quaternion $\mathbb{H}\in\,\mathbb{Q}$
can be written as
\begin{align}
\mathbb{H}\,\,= & \,\,e_{0}w+e_{1}x+e_{2}y+e_{3}z\,=\,\,e_{0}w+\sum_{j=1}^{3}e_{j}r_{j}\,,\label{eq:10}
\end{align}
where $w,x,y,z$ are the real numbers. The quaternion may also be
composed of the scalar and vector parts as
\begin{align}
\mathbb{H}\,\,= & \,\,S_{\mathbb{H}}+V_{\mathbb{H}}\,.\label{eq:11}
\end{align}
Here $e_{0}w$ is the scalar part of quaternion denoted by $S_{\mathbb{H}}$
and $\left(e_{1}x+e_{2}y+e_{3}z\right)$ is the vector part of quaternion
denoted by $V_{\mathbb{H}}.$ If scalar part is zero in equation (\ref{eq:11}),
then
\begin{align}
\mathbb{H}\,:\rightarrow\,\,V_{\mathbb{H}}\,\,=\,\, & e_{1}x+e_{2}y+e_{3}z\,,\label{eq:12}
\end{align}
is known as right quaternion or pure quaternion. Although any quaternion
can be seen as a vector in a four-dimensional vector space, it is
usually referred to as the pure quaternion as a vector. Conditionally,
if the vector part is zero then
\begin{align}
\mathbb{H}\,:\rightarrow\,\,S_{\mathbb{H}}\,\,=\,\, & e_{0}w\,,\label{eq:13}
\end{align}
is known as scalar quaternion. The addition of two quaternions will
be
\begin{align}
\mathbb{A+B\,\,=} & \,\,\left(e_{0}A_{0}+e_{1}A_{1}+e_{2}A_{2}+e_{3}A_{3}\right)+\left(e_{0}B_{0}+e_{1}B_{1}+e_{2}B_{2}+e_{3}B_{3}\right)\nonumber \\
\mathbb{=} & \,\,e_{0}\left(A_{0}+B_{0}\right)+e_{1}\left(A_{1}+B_{1}\right)+e_{2}\left(A_{2}+B_{2}\right)+e_{3}\left(A_{3}+B_{3}\right)\nonumber \\
= & \,\,e_{0}\left(H_{0}\right)+e_{j}\left(H_{j}\right)\,\,\,\,\,\forall\,\,\,(j=1,2,3)\,,\label{eq:14}
\end{align}
where $H_{0}\sim(A_{0}+B_{0})$ is the scalar part of quaternion while
$H_{j}\sim(A_{j}+B_{j})$ is the vector part of quaternion. Therefore,
the equation (\ref{eq:14}) shows the closure property concerning
quaternionic addition. The quaternionic algebra also satisfies the
associative and commutative properties of addition. Further, the multiplication
properties of two quaternions can be expressed by
\begin{align}
\mathbb{A\circ B}\,\,=\,\, & \left(e_{0}A_{0}+e_{1}A_{1}+e_{2}A_{2}+e_{3}A_{3}\right)\circ\left(e_{0}B_{0}+e_{1}B_{1}+e_{2}B_{2}+e_{3}B_{3}\right)\nonumber \\
=\,\, & e_{0}P_{0}+e_{1}P_{1}+e_{2}P_{2}+e_{3}P_{3}\,\,=\,\,\mathbb{P}\in\,\mathbb{Q}\,,\label{eq:15}
\end{align}
where $'\circ'$ is a symbol used for the quaternionic multiplication
and the components of $\mathbb{P}$ given in equation (\ref{eq:15})
are written by
\begin{align}
P_{0}:=\,\, & \left(A_{0}B_{0}-A_{1}B_{1}-A_{2}B_{2}-A_{3}B_{3}\right)\,\,\,\,\,\,\,\,\,\text{(coefficient of \ensuremath{e_{0}})}\nonumber \\
P_{1}:=\,\, & \left(A_{0}B_{1}+A_{1}B_{0}+A_{2}B_{3}-A_{3}B_{2}\right)\,\,\,\,\,\,\,\,\,(\text{coefficient of \ensuremath{e_{1}}})\nonumber \\
P_{2}:=\,\, & \left(A_{0}B_{2}+A_{2}B_{0}+A_{3}B_{1}-A_{1}B_{3}\right)\,\,\,\,\,\,\,\,\,(\text{coefficient of \ensuremath{e_{2}}})\label{eq:16}\\
P_{3}:=\,\, & \left(A_{0}B_{3}+A_{3}B_{0}+A_{1}B_{2}-A_{2}B_{1}\right)\,\,\,\,\,\,\,\,\,\,(\text{coefficient of \ensuremath{e_{3}})}.\nonumber 
\end{align}
Here the quaternionic multiplication follows the given rules for basis
$\left(e_{0},e_{1},e_{2},e_{3}\right),$ i.e.,
\begin{align}
e_{0}^{2}\,\,= & \,\,1\,,\,\,\,\,\,e_{i}^{2}\,\,=\,\,-1\,,\,\,\,e_{0}e_{i}\,\,=\,\,e_{i}e_{0}=\,\,e_{i}\,,\nonumber \\
e_{i}e_{j}\,\,= & \,\,-\delta_{ij}e_{0}+\epsilon_{ijk}e_{k}\,,\,\,\,\,\,\forall\,\,\,(i,j,k=\,\,1,2,3)\,.\label{eq:17}
\end{align}
All indices in Levi-Civita symbol $\epsilon_{ijk}$ are antisymmetric
and satisfy $e_{i}\times e_{j}=\,\,\sum_{k}\epsilon_{ijk}e_{k}.$
Thus, equation (\ref{eq:15}) can be written in compact form in terms
of ordinary dot and cross product as,
\begin{align}
\mathbb{A\circ B\,\,=} & \,\,e_{0}\left(A_{0}B_{0}-\overrightarrow{A}\cdot\overrightarrow{B}\right)+e_{j}\left[A_{0}\overrightarrow{B}+B_{0}\overrightarrow{A}+\left(\overrightarrow{A}\times\overrightarrow{B}\right)_{j}\right]\,.\label{eq:18}
\end{align}
In order to check the non-commutative property, we may write
\begin{align}
\mathbb{B\circ A}\,\,= & \,\,\left(e_{0}B_{0}+e_{1}B_{1}+e_{2}B_{2}+e_{3}B_{3}\right)\circ\left(e_{0}A_{0}+e_{1}A_{1}+e_{2}A_{2}+e_{3}A_{3}\right)\nonumber \\
\mathbb{=} & \,\,e_{0}\left(B_{0}A_{0}-B_{1}A_{1}-B_{2}A_{2}-B_{3}A_{3}\right)+e_{1}\left(B_{0}A_{1}+B_{1}A_{0}+B_{2}A_{3}-B_{3}A_{2}\right)\nonumber \\
\, & +e_{2}\left(B_{0}A_{2}+B_{2}A_{0}+B_{3}A_{1}-B_{1}A_{3}\right)+e_{3}\left(B_{0}A_{3}+B_{3}A_{0}+B_{1}A_{2}-B_{2}A_{1}\right)\nonumber \\
= & \,\,e_{0}\left(B_{0}A_{0}-\overrightarrow{B}\cdot\overrightarrow{A}\right)+e_{j}\left[B_{0}\overrightarrow{A}+A_{0}\overrightarrow{B}+\left(\overrightarrow{B}\times\overrightarrow{A}\right)_{j}\right]\,.\nonumber \\
\neq & \,\,\mathbb{A\circ B}\,.\label{eq:19}
\end{align}
From equation (\ref{eq:19}) it is clear that the product of two quaternions
is non-commutative because the product of two vectors is always non-commutative,
so that $\overrightarrow{A}\times\overrightarrow{B}=-\overrightarrow{B}\times\overrightarrow{A}.$
On the other hand, the multiplication of quaternions satisfy the associative
property i.e., $\left(\mathbb{A}\circ\mathbb{B}\right)\circ\mathbb{C}=\,\mathbb{A}\circ\left(\mathbb{B}\circ\mathbb{C}\right)$.
Further, the quaternionic conjugate of equation (\ref{eq:10}) can
be expressed as
\begin{align}
\mathbb{H^{\ast}}\,\,=\,\, & e_{0}w-\left(e_{1}x+e_{2}y+e_{3}z\right)\,,\label{eq:20}
\end{align}
and the product of two quaternions is given by \cite{key-41}
\begin{align}
\mathbb{A\cdot B\,\,=}\,\, & -\frac{1}{2}\left(\mathbb{A\circ B^{\ast}+\mathbb{B\circ A^{\ast}}}\right)=\,\,-\frac{1}{2}\left(\mathbb{A^{\ast}\circ B+\mathbb{B^{\ast}\circ A}}\right)\,.\label{eq:21}
\end{align}
The norm of a quaternion $\mathbb{H}$ denoted by $|\mathbb{H}|$
can be represented as
\begin{align}
|\mathbb{H}|\,\,=\,\, & \sqrt{\mathbb{H}\circ\mathbb{H^{\ast}}}=\,\,\sqrt{w^{2}+x^{2}+y^{2}+z^{2}}\,.\label{eq:22}
\end{align}
Now, the inverse of a quaternion can be defined by
\begin{align}
\mathbb{H}^{-1}\,\,=\,\, & \frac{\mathbb{H^{\ast}}}{|\mathbb{H}|^{2}}\,\,\equiv\,\,\frac{e_{0}w-e_{1}x-e_{2}y-e_{3}z}{w^{2}+x^{2}+y^{2}+z^{2}}\,.\label{eq:23}
\end{align}
Moreover,\textbf{ }the quotient of two vectors is also known as quaternion
which can be represented by \cite{key-42}
\begin{align}
\mathbb{\mathbb{H}\,\,=} & \,\,\frac{\alpha}{\beta}\,,\label{eq:24}
\end{align}
where $\mathbf{\alpha}$ and $\beta$ are the two vectors. The another
approach on a ``Tensor of a quaternion'' $\left(\mathbf{T}_{\mathbb{H}}\right)$
is given by Hamilton \cite{key-42}, which can be written by
\begin{align}
\left(\mathbf{T_{\mathbb{H}}}\right)^{\mathrm{2}}\,\,=\,\, & \mathbb{H\circ H^{\ast}}\,\,=\,\,w^{2}+x^{2}+y^{2}+z^{2},\label{eq:25}
\end{align}
so that $T_{\mathbb{H}}=\sqrt{\left(w^{2}+x^{2}+y^{2}+z^{2}\right)}\,.$
The tensor of a quaternion is similar to the norm of a quaternion,
it will be always positive or real number \cite{key-43}. If the scalar
part is zero then $\mathbf{T}\left(\mathbf{V}_{\mathbb{H}}\right)\,=\,\sqrt{\left(x^{2}+y^{2}+z^{2}\right)}$
is known as the tensor of a vector quaternion and if vector part is
zero then $\mathbf{T\left(\mathbf{S_{\mathbb{H}}}\right)}\,=\,w$
is known as the tensor of a scalar quaternion. Further, the 'versor'
of a quaternion which indicates direction, is a unit quaternion known
as a normalized quaternion $(\mathbf{U_{\mathbb{H}}})$, so that
\begin{align}
\mathbf{U_{\mathbb{H}}\,\,=\,\,} & \frac{\mathbb{H}}{|\mathbb{H}|}\,\,\equiv\,\,\frac{e_{0}w+e_{1}x+e_{2}y+e_{3}z}{\sqrt{\left(w^{2}+x^{2}+y^{2}+z^{2}\right)}}\,.\label{eq:26}
\end{align}
The scalar and vector parts of a quaternionic versor can be expressed
by $\mathbf{\mathbf{S}\left(\mathbf{U_{\mathbb{H}}}\right)\,\,=}$
$w\left(w^{2}+x^{2}+y^{2}+z^{2}\right)^{-1/2}$ and $\mathbf{V\left(\mathbf{U_{\mathbb{H}}}\right)\,\,=\,\,}(e_{1}x+e_{2}y+e_{3}z)$
$\left(w^{2}+x^{2}+y^{2}+z^{2}\right)^{-1/2}$. A quaternion can also
be written in terms of tensor of a quaternion and versor of a quaternion
as \cite{key-42}
\begin{align}
\mathbb{H}\,\,=\,\, & \mathbf{T}_{\mathbb{H}}\mathbf{U}_{\mathbb{H}}=\,\,e_{0}w+e_{1}x+e_{2}y+e_{3}z\,.\label{eq:27}
\end{align}

\section{Transformation of quaternionic basis elements}

If a system of coordinates rotates from its original frame of reference
then there will be a change in unit vectors to original unit vectors.
This change of unit vectors represents the transformation equations.
Here, we introduce the transformation equations for the 2-dimensional
system in which the equations are used for the transformation of coordinates
from one plane to another plane, i.e.,
\begin{align}
x^{'}\,\,=\,\,x\cos\phi+y\sin\phi\,,\,\,\,\,\,\,\,\,\,y^{'}\,\,=\,\, & -x\sin\phi+y\cos\phi\,,\label{eq:28}
\end{align}
which gives
\begin{align}
\frac{\partial x^{'}}{\partial x}\,\,=\,\, & \cos\phi,\,\,\,\,\,\frac{\partial x^{'}}{\partial y}\,\,=\,\,\sin\phi,\,\,\,\,\,\frac{\partial y^{'}}{\partial x}\,\,=\,\,-\sin\phi,\,\,\,\,\,\frac{\partial y^{'}}{\partial y}\,\,=\,\,\cos\phi\,.\label{eq:29}
\end{align}
Thus, the transformation of basis elements can be written in the matrix
form
\begin{align}
\left(\begin{array}{c}
e_{1}^{'}\\
e_{2}^{'}
\end{array}\right)\,\,:\longmapsto\,\, & \left(\begin{array}{cc}
\cos\phi & \sin\phi\\
-\sin\phi & \cos\phi
\end{array}\right)\left(\begin{array}{c}
e_{1}\\
e_{2}
\end{array}\right)\,,\label{eq:30}
\end{align}
which leads to following linear transformation equations
\begin{align}
e_{1}^{'}\,\longmapsto\,\, & \frac{\partial x^{'}}{\partial x}e_{1}+\frac{\partial x^{'}}{\partial y}e_{2}\,,\nonumber \\
e_{2}^{'}\,\longmapsto\,\, & \frac{\partial y^{'}}{\partial x}e_{1}+\frac{\partial y^{'}}{\partial y}e_{2}\,,\label{eq:31}
\end{align}
where $e_{1}$ and $e_{2}$ are the unit vectors corresponding to
$x\,\text{and}\,y-$ axis in $XY-$ plane, while $e_{1}^{'}$ and
$e_{2}^{'}$ are the unit vectors corresponding to $x^{'}\,\text{and}\,y^{'}-$
axis in $X^{'}Y^{'}-$ plane, respectively. $\phi$ is the angle between
these two planes. Like 2-D transformation, we may extend the transformation
relations of basis elements for 3-D system. Thus, the 3-D transformation
matrix $\left(D\right)$ can be written by
\begin{align}
D\,\,=\,\, & \left(\begin{array}{ccc}
\cos\psi\cos\phi-\sin\psi\sin\phi\cos\theta & -\cos\psi\sin\phi-\sin\psi\cos\theta\cos\phi & \sin\psi\sin\theta\\
\sin\psi\cos\phi+\cos\psi\sin\phi\cos\theta & -\sin\psi\sin\phi+\cos\psi\cos\theta\cos\phi & -\cos\psi\sin\theta\\
\sin\theta\sin\phi & \sin\theta\cos\phi & \cos\theta
\end{array}\right)\,,\label{eq:32}
\end{align}
where $\psi,\theta,\phi$ are the three independent parameters known
as \textit{Euler angles} which represents the rotation of axis in
new frame of reference with respect to original axis. Let $\left(e_{1},e_{2},e_{3}\right)$
are pure quaternionic unit vectors along $x,y,z$- axis in $S-$ frame
while $\left(e_{1}^{'},e_{2}^{'},e_{3}^{'}\right)$ are unit vectors
for $x^{'},y^{'},z^{'}$- axis in $S^{'}-$ frame, then the linear
transformation relations for pure quaternionic unit vectors become,
\begin{align}
e_{1}^{'}\,\longmapsto & \,\,\frac{\partial x^{'}}{\partial x}e_{1}+\frac{\partial x^{'}}{\partial y}e_{2}+\frac{\partial x^{'}}{\partial z}e_{3}\,,\nonumber \\
e_{2}^{'}\,\longmapsto & \,\,\frac{\partial y^{'}}{\partial x}e_{1}+\frac{\partial y^{'}}{\partial y}e_{2}+\frac{\partial y^{'}}{\partial z}e_{3}\,,\nonumber \\
e_{3}^{'}\,\longmapsto & \,\,\frac{\partial z^{'}}{\partial x}e_{1}+\frac{\partial z^{'}}{\partial y}e_{2}+\frac{\partial z^{'}}{\partial z}e_{3}\,,\label{eq:33}
\end{align}
where the derivatives become
\begin{align*}
\frac{\partial x^{'}}{\partial x}\,=\,\, & \cos\psi\cos\phi-\sin\psi\sin\phi\cos\theta,\,\,\,\,\,\,\,\,\,\,\,\,\,\,\frac{\partial x^{'}}{\partial y}\,=\,\,-\cos\psi\sin\phi-\sin\psi\cos\theta\cos\phi\,,\\
\frac{\partial x^{'}}{\partial z}\,=\,\, & \sin\psi\sin\theta,\,\,\,\,\,\,\,\,\,\,\,\,\,\,\,\frac{\partial y^{'}}{\partial x}\,=\,\,\sin\psi\cos\phi+\cos\psi\sin\phi\cos\theta\,,\\
\frac{\partial y^{'}}{\partial y}\,=\,\, & -\sin\psi\sin\phi+\cos\psi\cos\theta\cos\phi,\,\,\,\,\,\,\,\,\,\,\,\,\,\,\,\frac{\partial y^{'}}{\partial z}\,=\,\,-\cos\psi\sin\theta\,,\\
\frac{\partial z^{'}}{\partial x}\,=\,\, & \sin\theta\sin\phi,\,\,\,\,\,\,\,\,\,\,\,\,\,\,\,\,\,\,\,\frac{\partial z^{'}}{\partial y}\,=\,\,\sin\theta\cos\phi,\,\,\,\,\,\,\,\,\,\,\,\,\,\,\,\,\,\,\frac{\partial z^{'}}{\partial z}\,=\,\,\cos\theta\,.
\end{align*}
Interestingly, these transformation equations satisfy the properties
of vector algebra. The cross product of two vectors is always non-commutative
for pure quaternionic space. Moreover, the properties of pure quaternionic
unit elements $\left(e_{1},e_{2},e_{3}\right)$ corresponding to vector
algebra basis can be represented as
\begin{align}
e_{1}^{'}\circ e_{1}^{'}\,=\,\, & \left(\frac{\partial x^{'}}{\partial x}e_{1}+\frac{\partial x^{'}}{\partial y}e_{2}+\frac{\partial x^{'}}{\partial z}e_{3}\right)\circ\left(\frac{\partial x^{'}}{\partial x}e_{1}+\frac{\partial x^{'}}{\partial y}e_{2}+\frac{\partial x^{'}}{\partial z}e_{3}\right)\nonumber \\
=\,\, & -\left(\frac{\partial x^{'}}{\partial x}\right)^{2}-\left(\frac{\partial x^{'}}{\partial y}\right)^{2}-\left(\frac{\partial x^{'}}{\partial z}\right)^{2}\nonumber \\
=\,\, & -\left(\cos\psi\cos\phi-\sin\psi\sin\phi\cos\theta\right)^{2}-\left(-\cos\psi\sin\phi-\sin\psi\cos\theta\cos\phi\right)^{2}\nonumber \\
\,\, & -\left(\sin\psi\sin\theta\right)^{2}=\,\,-1\,.\label{eq:34}
\end{align}
and
\begin{align}
e_{2}^{'}\circ e_{2}^{'}\,\,=\,\, & -1,\,\,\,\,\,\,\,e_{3}^{'}\circ e_{3}^{'}\,\,=\,\,-1\,.\label{eq:35}
\end{align}
On the other hand
\begin{align*}
e_{1}^{'}\circ e_{2}^{'}\,\,=\,\, & \left(\frac{\partial x^{'}}{\partial x}e_{1}+\frac{\partial x^{'}}{\partial y}e_{2}+\frac{\partial x^{'}}{\partial z}e_{3}\right)\circ\left(\frac{\partial y^{'}}{\partial x}e_{1}+\frac{\partial y^{'}}{\partial y}e_{2}+\frac{\partial y^{'}}{\partial z}e_{3}\right)\\
=\,\, & -\frac{\partial x^{'}}{\partial x}\frac{\partial y^{'}}{\partial x}-\frac{\partial x^{'}}{\partial y}\frac{\partial y^{'}}{\partial y}-\frac{\partial x^{'}}{\partial z}\frac{\partial y^{'}}{\partial z}-\left(\frac{\partial x^{'}}{\partial z}\frac{\partial y^{'}}{\partial y}-\frac{\partial x^{'}}{\partial y}\frac{\partial y^{'}}{\partial z}\right)e_{1}\\
-\,\, & \left(\frac{\partial x^{'}}{\partial x}\frac{\partial y^{'}}{\partial z}-\frac{\partial x^{'}}{\partial z}\frac{\partial y^{'}}{\partial x}\right)e_{2}-\left(\frac{\partial x^{'}}{\partial y}\frac{\partial y^{'}}{\partial x}-\frac{\partial x^{'}}{\partial x}\frac{\partial y^{'}}{\partial y}\right)e_{3}
\end{align*}
\begin{align}
=\,\, & -\left(\cos\psi\cos\phi-\sin\psi\sin\phi\cos\theta\right)\left(\sin\psi\cos\phi+\cos\psi\sin\phi\cos\theta\right)\nonumber \\
-\,\, & \left(-\cos\psi\sin\phi-\sin\psi\cos\theta\cos\phi\right)\left(-\sin\psi\sin\phi+\cos\psi\cos\theta\cos\phi\right)\nonumber \\
-\,\, & \left(\sin\psi\sin\theta\right)\left(-\cos\psi\sin\theta\right)-e_{1}[\left(\sin\psi\sin\theta\right)\left(-\sin\psi\sin\phi+\cos\psi\cos\theta\cos\phi\right)\nonumber \\
+\,\, & \left(-\cos\psi\sin\phi-\sin\psi\cos\theta\cos\phi\right)\left(-\cos\psi\sin\theta\right)]\nonumber \\
-\,\, & e_{2}[\left(\cos\psi\cos\phi-\sin\psi\sin\phi\cos\theta\right)\left(-\cos\psi\sin\theta\right)\nonumber \\
+\,\, & \left(\sin\psi\sin\theta\right)\left(\sin\psi\cos\phi+\cos\psi\sin\phi\cos\theta\right)]\nonumber \\
-\,\, & e_{3}[\left(-\cos\psi\sin\phi-\sin\psi\cos\theta\cos\phi\right)\left(\sin\psi\cos\phi+\cos\psi\sin\phi\cos\theta\right)\nonumber \\
+\,\, & \left(\cos\psi\cos\phi-\sin\psi\sin\phi\cos\theta\right)\left(-\sin\psi\sin\phi+\cos\psi\cos\theta\cos\phi\right)]\nonumber \\
=\,\, & e_{3}^{'}\,,\label{eq:36}
\end{align}
and
\begin{align}
e_{2}^{'}\circ e_{1}^{'}\,\,=\,\, & \left(\frac{\partial y^{'}}{\partial x}e_{1}+\frac{\partial y^{'}}{\partial y}e_{2}+\frac{\partial y^{'}}{\partial z}e_{3}\right)\circ\left(\frac{\partial x^{'}}{\partial x}e_{1}+\frac{\partial x^{'}}{\partial y}e_{2}+\frac{\partial x^{'}}{\partial z}e_{3}\right)\nonumber \\
=\,\, & -\frac{\partial y^{'}}{\partial x}\frac{\partial x^{'}}{\partial x}-\frac{\partial y^{'}}{\partial y}\frac{\partial x^{'}}{\partial y}-\frac{\partial y^{'}}{\partial z}\frac{\partial x^{'}}{\partial z}+\left(\frac{\partial y^{'}}{\partial y}\frac{\partial x^{'}}{\partial z}-\frac{\partial y^{'}}{\partial z}\frac{\partial x^{'}}{\partial y}\right)e_{1}\nonumber \\
+\,\, & \left(\frac{\partial y^{'}}{\partial z}\frac{\partial x^{'}}{\partial x}-\frac{\partial y^{'}}{\partial x}\frac{\partial x^{'}}{\partial z}\right)e_{2}+\left(\frac{\partial y^{'}}{\partial x}\frac{\partial x^{'}}{\partial y}-\frac{\partial y^{'}}{\partial y}\frac{\partial x^{'}}{\partial x}\right)e_{3}\nonumber \\
=\,\, & -\left(\sin\psi\cos\phi+\cos\psi\sin\phi\cos\theta\right)\left(\cos\psi\cos\phi-\sin\psi\sin\phi\cos\theta\right)\nonumber \\
-\,\, & \left(-\sin\psi\sin\phi+\cos\psi\cos\theta\cos\phi\right)\left(-\cos\psi\sin\phi-\sin\psi\cos\theta\cos\phi\right)\nonumber \\
-\,\, & \left(-\cos\psi\sin\theta\right)\left(\sin\psi\sin\theta\right)+e_{1}[\left(\sin\psi\sin\theta\right)\left(-\sin\psi\sin\phi+\cos\psi\cos\theta\cos\phi\right)\nonumber \\
-\,\, & \left(-\cos\psi\sin\phi-\sin\psi\cos\theta\cos\phi\right)\left(-\cos\psi\sin\theta\right)]\nonumber \\
+\,\, & e_{2}[\left(\cos\psi\cos\phi-\sin\psi\sin\phi\cos\theta\right)\left(-\cos\psi\sin\theta\right)\nonumber \\
-\,\, & \left(\sin\psi\sin\theta\right)\left(\sin\psi\cos\phi+\cos\psi\sin\phi\cos\theta\right)]\nonumber \\
+\,\, & e_{3}[\left(-\cos\psi\sin\phi-\sin\psi\cos\theta\cos\phi\right)\left(\sin\psi\cos\phi+\cos\psi\sin\phi\cos\theta\right)\nonumber \\
-\,\, & \left(\cos\psi\cos\phi-\sin\psi\sin\phi\cos\theta\right)\left(-\sin\psi\sin\phi+\cos\psi\cos\theta\cos\phi\right)]\nonumber \\
=\,\,\, & -e_{3}^{'}\,.\label{eq:37}
\end{align}
Similarly, we obtain
\begin{align}
e_{2}^{'}\circ e_{3}^{'}\,\,=\,\, & e_{1}^{'}\,,\,\,\,\,e_{3}^{'}\circ e_{2}^{'}\,=\,\,-e_{1}^{'}\,,\nonumber \\
e_{3}^{'}\circ e_{1}^{'}\,\,=\,\, & e_{2}^{'}\,,\,\,\,\,e_{1}^{'}\circ e_{3}^{'}\,=\,\,-e_{2}^{'}\,.\label{eq:38}
\end{align}
Now, we easily can extend the pure-quaternion to quaternion by adding
the scalar unit element $e_{0}$. It should be noticed that the scalar
unit element becomes invariant under transformation given in equation
(\ref{eq:6}) but it plays an important role with unit elements $e_{1},e_{2},e_{3}$
in quaternionic transformation.\textbf{ }Myszkowski\textbf{ }\cite{key-44}
has given the idea of 4-D transformation similar to the 3-D transformation.
Thus, we assume the transformation equations for quaternionic basis
elements as
\begin{align}
e_{0}^{'}\,\longmapsto\,\, & e_{0}\,,\nonumber \\
e_{1}^{'}\,\longmapsto\,\, & \frac{\partial x^{'}}{\partial t}e_{0}+\frac{\partial x^{'}}{\partial x}e_{1}+\frac{\partial x^{'}}{\partial y}e_{2}+\frac{\partial x^{'}}{\partial z}e_{3}\,,\nonumber \\
e_{2}^{'}\,\longmapsto\,\, & \frac{\partial y^{'}}{\partial t}e_{0}+\frac{\partial y^{'}}{\partial x}e_{1}+\frac{\partial y^{'}}{\partial y}e_{2}+\frac{\partial y^{'}}{\partial z}e_{3}\,,\nonumber \\
e_{3}^{'}\,\longmapsto\,\, & \frac{\partial z^{'}}{\partial t}e_{0}+\frac{\partial z^{'}}{\partial x}e_{1}+\frac{\partial z^{'}}{\partial y}e_{2}+\frac{\partial z^{'}}{\partial z}e_{3}\,.\label{eq:39}
\end{align}
Here, we focused on space-time structure for quaternionic transformation,
the quaternionic properties will also be satisfied as similar to the
3-D rotation. Now, in the next section, we shall interpret these relations
for Riemannian geometry.

\section{Quaternionic approach on Riemannian geometry}

\subsection{Quaternion transformation}

In Riemannian geometry, the motion of object takes place in curvature
space-time in which 4-dimensional frame of reference rotate along
the curve. The structure of quaternionic space-time coordinates can
be written as $P^{\mu}\equiv(P^{0},P^{j})\,\simeq\,\,\left(P^{\xi},P^{1},P^{2},P^{3}\right)$.
The time coordinate may refer corresponding to $e_{0}$ i.e. $t\mapsto P^{\xi},$
and the spatial coordinates may refer corresponding to $e_{j}$ i.e.
$(x,y,z)\mapsto\left(P^{1},P^{2},P^{3}\right)$ where $j=\,\,1,2,3$.
Now, the quaternionic elements may transformation as
\begin{align}
e_{0}^{'}\,\,=\,\, & e_{0},\,\,\,\,e_{i}^{'}\,\,=\,\,\frac{\partial P^{'i}}{\partial P^{\mu}}e_{\mu},\,\,\:\,\,\,\,\,\,\,\,\,\,\,\,\,\,\forall\,\,(i=\,\,1,2,3)\,.\label{eq:40}
\end{align}
The transformation of scalar and vector field components of a quaternion
variable ($\mathbb{H}$) from $S$ to $S^{'}$-frame can be expressed
as
\begin{align}
H^{\xi}\,\,=\,\, & H^{'\xi}+\frac{\partial P^{\xi}}{\partial P^{'i}}H^{'i}\,\,\,\,\,\,\,\,\,\,\,\,\,\,\,\,\,\,\,\,\,\text{(coefficient of }e_{0}\text{)}\,,\label{eq:41}\\
H^{j}\,\,=\,\, & \frac{\partial P^{j}}{\partial P^{'i}}H^{'i}\,\,\,\,\,\,\,\,\,\,\,\,\,\,\,\,\,\,\,\,\,\,\,\,\,\,\,\,\,\,\,\,\,\,\,\,\,\,\,\text{(coefficient of }e_{j}\text{)}\,.\label{eq:42}
\end{align}
In the above transformation relation (\ref{eq:41}), we assume that
the effect of $H^{'\xi}$ on $H^{\xi}$ is linear because it shows
the simple transformation on flat space-time. So, we can neglect the
component $H^{'\xi}$ to get the quaternionic transformation (Q-transformation)
in curved space-time. Therefore, we have
\begin{align}
\mathbb{H}^{'}\,\,:\longmapsto\,\,\mathbb{H}\,\,=\,\,\, & \left(\frac{\partial P^{\xi}}{\partial P^{'i}}H^{'i},\,\,\frac{\partial P^{j}}{\partial P^{'i}}H^{'i}\right)\,.\label{eq:43}
\end{align}
Equation (\ref{eq:43}) represents the Q-transformation in contravariant
form, we can also write the above transformation relations in quaternionic
covariant form as
\begin{align}
H_{\xi}\,\,=\,\, & \frac{\partial P^{'i}}{\partial P^{\xi}}H_{i}^{'}\,\,,\,\,\,\,\,\,H_{j}\,\,=\,\,\frac{\partial P^{'i}}{\partial P^{j}}H_{i}^{'}\,\,.\label{eq:44}
\end{align}
Now, we can write the transformation of quaternionic differential
operator $\square\simeq\{\partial_{\phi},\,\partial_{j}\}$ as
\begin{align}
\partial_{\phi}\,\,=\,\, & \frac{\partial P^{\phi}}{\partial P^{'k}}\partial_{k}^{'}\,,\,\,\,\,\,\,\,\,\,\,\partial_{j}\,\,=\,\,\frac{\partial P^{j}}{\partial P^{'k}}\partial_{k}^{'}\,,\,\,\,\,\,\,\,\,\,\,\,\,\forall\,(j,k=1,2,3)\,.\label{eq:45}
\end{align}
Since the components of a quaternion can also represent in form of
coordinates, then we can write the transformation of quaternionic
coordinates\textbf{ }\cite{key-36},
\begin{align}
dP^{\xi}\,\,=\,\, & \frac{\partial P^{\xi}}{\partial P^{'i}}\,dP^{'i}\,,\,\,\,\,\,\,\,\,\,dP^{j}\,\,=\,\,\frac{\partial P^{j}}{\partial P^{'i}}\,dP^{'i},\,\,\,\,\,\,\,\forall\,\left(i,j=\,\,1,2,3\right)\,.\label{eq:46}
\end{align}

\begin{onehalfspace}

\subsection{Quaternionic covariant derivative}
\end{onehalfspace}

In the quaternionic form of curvature space-time, the scalar field
derivative of equation (\ref{eq:44}) can be written as
\begin{align}
H_{;\phi}^{\xi}\,\,=\,\, & H_{,\phi}^{\xi}+\left\{ \begin{array}{c}
\xi\\
\phi\phi
\end{array}\right\} \,H^{\phi}\,,\label{eq:47}
\end{align}
where the Christoffel symbol is denoted by $\left\{ \begin{array}{c}
\xi\\
\phi\phi
\end{array}\right\} \,=\,\frac{\partial}{\partial P^{\phi}}\left(\frac{\partial P^{\xi}}{\partial P^{\phi}}\right)$. Similarly, we have
\begin{align}
H_{\xi;\phi}\,\,=\,\, & H_{\xi,\phi}-\left\{ \begin{array}{c}
\phi\\
\phi\xi
\end{array}\right\} \,H_{\phi}\,.\label{eq:48}
\end{align}
Here the role of Christoffel symbol is very important for Riemannian
space-time. If $\left\{ \begin{array}{c}
\xi\\
\phi\phi
\end{array}\right\} \rightarrow\,0$, then the quaternionic curvature covariant derivative transforms
into the quaternionic usual partial derivative which indicates the
linear transformation. Correspondingly, applying quaternionic derivative
$\partial_{\phi}$ on $H^{j}$, we obtain
\begin{align*}
H_{;\phi}^{j}\,\,=\,\, & H_{,\phi}^{j}+\left\{ \begin{array}{c}
j\\
\phi\phi
\end{array}\right\} \,H^{\phi}\,.
\end{align*}
As such, we also can operate the quaternionic derivative $\partial_{k}$
($k=1,2,3)$ on given equation (\ref{eq:44}), so that
\begin{align}
H_{;k}^{\xi}\,\,=\,\, & H_{,k}^{\xi}+\left\{ \begin{array}{c}
\xi\\
km
\end{array}\right\} \,H^{m}\,,\label{eq:49}\\
H_{;k}^{j}\,\,=\,\, & H_{,k}^{j}+\left\{ \begin{array}{c}
j\\
km
\end{array}\right\} \,H^{m}\,,\label{eq:50}
\end{align}
where $j,k,m=1,2,3.$ Equations (\ref{eq:49}) and (\ref{eq:50})
are represented scalar to vector and vector to vector field transformations,
respectively, for generalized quaternionic fields \cite{key-42}.
It should be remarked that through quaternionic derivative, the generalized
covariant derivative gives enlargement to the rank of quaternionic
tensors which is important to provide more information about the transformed
coordinate. In more general the quaternionic tensor derivatives are
\begin{align}
T_{\xi\xi;\phi}\,\,=\,\, & T_{\xi\xi,\phi}-\,\left\{ \begin{array}{c}
\eta\\
\phi\xi
\end{array}\right\} T_{\xi\eta}-\,\left\{ \begin{array}{c}
\eta\\
\phi\xi
\end{array}\right\} T_{\eta\xi}\,,\label{eq:51}\\
T_{\xi\xi;k}\,\,=\,\, & T_{\xi\xi,k}-\,\left\{ \begin{array}{c}
l\\
k\xi
\end{array}\right\} T_{\xi l}-\,\left\{ \begin{array}{c}
l\\
k\xi
\end{array}\right\} T_{l\xi}\,,\label{eq:52}\\
T_{\xi n;\phi}\,\,=\,\, & T_{\xi n,\phi}-\,\left\{ \begin{array}{c}
\eta\\
\phi n
\end{array}\right\} T_{\xi\eta}-\,\left\{ \begin{array}{c}
\eta\\
\phi\xi
\end{array}\right\} T_{\eta n}\,,\label{eq:53}\\
T_{\xi n;k}\,\,=\,\, & T_{\xi n,k}-\,\left\{ \begin{array}{c}
l\\
kn
\end{array}\right\} T_{\xi l}-\,\left\{ \begin{array}{c}
l\\
k\xi
\end{array}\right\} T_{ln}\,,\label{eq:54}\\
T_{j\xi;\phi}\,\,=\,\, & T_{j\xi,\phi}-\,\left\{ \begin{array}{c}
\eta\\
\phi\xi
\end{array}\right\} T_{j\eta}-\,\left\{ \begin{array}{c}
\eta\\
\phi j
\end{array}\right\} T_{\eta\xi}\,,\label{eq:55}\\
T_{j\xi;k}\,\,=\,\, & T_{j\xi,k}-\,\left\{ \begin{array}{c}
l\\
k\xi
\end{array}\right\} T_{jl}-\,\left\{ \begin{array}{c}
l\\
kj
\end{array}\right\} T_{l\xi}\,,\label{eq:56}\\
T_{jn;\phi}\,\,=\,\, & T_{jn,\phi}-\,\left\{ \begin{array}{c}
\eta\\
\phi n
\end{array}\right\} T_{j\eta}-\,\left\{ \begin{array}{c}
\eta\\
\phi j
\end{array}\right\} T_{\eta n}\,,\label{eq:57}\\
T_{jn;k}\,\,=\,\, & T_{jn,k}-\,\left\{ \begin{array}{c}
l\\
kn
\end{array}\right\} T_{jl}-\,\left\{ \begin{array}{c}
l\\
kj
\end{array}\right\} T_{ln}\,,\label{eq:58}
\end{align}
where the right-hand side of equations (\ref{eq:51})-(\ref{eq:58}),
the first term shows the tensorial transformation of derivative of
quaternionic tensor and the last two terms show the Christoffel symbols
which tell us how the geodesic path changes from point to point.

\subsection{Quaternionic metric tensor}

To expressing the quaternionic metric tensor in terms of the Christoffel
symbol, we can use the covariant derivative of quaternionic tensor.
Therefore, one can write
\begin{align}
g_{\xi\xi;\phi}\,\,=\,\, & g_{\xi\xi,\phi}-\,\left\{ \begin{array}{c}
\eta\\
\phi\xi
\end{array}\right\} g_{\xi\eta}-\,\left\{ \begin{array}{c}
\eta\\
\phi\xi
\end{array}\right\} g_{\eta\xi}\,.\label{eq:59}
\end{align}
where the line element becomes $\mathsf{ds}=\sqrt{-g_{\mu\nu}(x)\,dx^{\mu}dx^{\nu}.}$
Under the covariant differentiation, the quaternionic metric tensor
is constant i.e.,
\begin{align}
0\,\,=\,\, & g_{\xi\xi,\phi}-\,\left\{ \begin{array}{c}
\eta\\
\phi\xi
\end{array}\right\} g_{\xi\eta}-\,\left\{ \begin{array}{c}
\eta\\
\phi\xi
\end{array}\right\} g_{\eta\xi}\nonumber \\
g_{\xi\xi,\phi}\,\,=\,\, & \left\{ \begin{array}{c}
\eta\\
\phi\xi
\end{array}\right\} g_{\xi\eta}+\,\left\{ \begin{array}{c}
\eta\\
\phi\xi
\end{array}\right\} g_{\eta\xi}\,.\label{eq:60}
\end{align}
Here the metric tensor $g_{im}^{'}=\,\,\frac{\partial P^{\mu}}{\partial P^{'i}}\,\frac{\partial P^{\nu}}{\partial P^{'m}}\,\Omega_{\mu\nu}$,
where $\Omega_{\mu\nu}$ is the four-dimensional Minkowski metric
define by
\begin{align*}
\Omega_{\mu\nu}\,\,=\,\, & \left(\begin{array}{cccc}
-1 & 0 & 0 & 0\\
0 & 1 & 0 & 0\\
0 & 0 & 1 & 0\\
0 & 0 & 0 & 1
\end{array}\right)\,,
\end{align*}
and the Christoffel symbol satisfy the following propertie\textbf{s
\begin{align}
\left\{ \begin{array}{c}
i\\
kl
\end{array}\right\} \,\,=\,\,\left\{ \begin{array}{c}
i\\
lk
\end{array}\right\} ,\,\,\,\,\,\,\,\left\{ \begin{array}{c}
m\\
il
\end{array}\right\} \,g_{mk}\,\,=\,\, & \left\{ \begin{array}{c}
k\\
il
\end{array}\right\} \,.\label{eq:61}
\end{align}
}Now, the value of Christoffel symbol corresponding to quaternionic
scalar field can be written as,
\begin{align}
\left\{ \begin{array}{c}
\eta\\
\xi\xi
\end{array}\right\} \,\,=\,\, & \frac{1}{2}g^{\eta\phi}\,\left(g_{\phi\xi,\xi}+g_{\xi\phi,\xi}-g_{\xi\xi,\phi}\right)\,,\label{eq:62}
\end{align}
where $g^{\eta\phi}$ is the inverse matrix of $g_{\eta\phi}$. Correspondingly,
in quaternionic pure vectorial field, the Christoffel symbol can be
played by
\begin{align}
\left\{ \begin{array}{c}
l\\
jn
\end{array}\right\} \,\,=\,\, & \frac{1}{2}g^{lk}\,\left(g_{kj,n}+g_{nk,j}-g_{jn,k}\right)\,,\label{eq:63}
\end{align}
and the other diversified quaternionic scalar and vector fields show
a mixed form of quaternionic Christoffel symbol in terms of the metric
tensor, i.e.
\begin{align}
\left\{ \begin{array}{c}
l\\
\xi\xi
\end{array}\right\} \,\,=\,\, & \frac{1}{2}g^{lk}\,\left(g_{k\xi,\xi}+g_{\xi k,\xi}-g_{\xi\xi,k}\right)\,,\label{eq:64}\\
\left\{ \begin{array}{c}
\eta\\
\xi n
\end{array}\right\} \,\,=\,\, & \frac{1}{2}g^{\eta\phi}\,\left(g_{\phi\xi,n}+g_{n\phi,\xi}-g_{\xi n,\phi}\right)\,,\label{eq:65}\\
\left\{ \begin{array}{c}
l\\
\xi n
\end{array}\right\} \,\,=\,\, & \frac{1}{2}g^{lk}\,\left(g_{k\xi,n}+g_{nk,\xi}-g_{\xi n,k}\right)\,,\label{eq:66}\\
\left\{ \begin{array}{c}
\eta\\
jn
\end{array}\right\} \,\,=\,\, & \frac{1}{2}g^{\eta\phi}\,\left(g_{\phi j,n}+g_{n\phi,j}-g_{jn,\phi}\right)\,,\label{eq:67}
\end{align}
where $(\xi,\eta,\phi)$ are used for generalized quaternionic scalar
field variables while $(i,j,k,l,m,n)$ are the generalized quaternionic
vector field variables.

\subsection{Quaternionic geodesic equation}

In general theory of relativity, geodesic equation gives the replacement
of linear space-time to curved space-time. In other words, a particle
moving in a curvature space-time is always follow the path of geodesic.
The geodesic path in curvature space-time structure can be described
in terms of four-vector form with the principle of least action, i.e.
$dH^{\mu}=\,0,$\textbf{ }see ref.\cite{key-3}. In the pure quaternionic
scalar field,
\begin{align}
H_{;\phi}^{\xi}\,\,\equiv\,\,H_{,\phi}^{\xi}+\left\{ \begin{array}{c}
\xi\\
\phi\phi
\end{array}\right\} \,H^{\phi}\,\,=\,\, & 0\nonumber \\
\Rightarrow\,dH^{\xi}+\left\{ \begin{array}{c}
\xi\\
\phi\phi
\end{array}\right\} \,dP^{\phi}\,H^{\phi}\,\,=\,\, & 0\,.\label{eq:68}
\end{align}
Now, dividing equation (\ref{eq:68}) by $ds$ (e.g. a scalar parameter
of motion as proper time) and substituting $H^{\xi}\,=\,\frac{dP^{\xi}}{ds}$,
$H^{\phi}\,=\,\frac{dP^{\phi}}{ds}$, we get
\begin{align}
\frac{d^{2}P^{\xi}}{ds^{2}}+\left\{ \begin{array}{c}
\xi\\
\phi\phi
\end{array}\right\} \,\frac{dP^{\phi}}{ds}\frac{dP^{\phi}}{ds}\,\,=\,\, & 0\,.\label{eq:69}
\end{align}
Similarly, for quaternionic vector fields, we get
\begin{align}
\frac{d^{2}P^{j}}{ds^{2}}+\left\{ \begin{array}{c}
j\\
km
\end{array}\right\} \,\frac{dP^{k}}{ds}\frac{dP^{m}}{ds}\,\,=\,\, & 0\,,\label{eq:70}
\end{align}
and for a mixed fields,
\begin{align}
\frac{d^{2}P^{j}}{ds^{2}}+\left\{ \begin{array}{c}
j\\
\phi\phi
\end{array}\right\} \,\frac{dP^{\phi}}{ds}\frac{dP^{\phi}}{ds}\,\,=\,\, & 0\,,\nonumber \\
\frac{d^{2}P^{\xi}}{ds^{2}}+\left\{ \begin{array}{c}
\xi\\
km
\end{array}\right\} \,\frac{dP^{k}}{ds}\frac{dP^{m}}{ds}\,\,=\,\, & 0\,.\label{eq:71}
\end{align}
Equations (\ref{eq:69})-(\ref{eq:71}) represent the non-linear equations
called the quaternionic geodesic equations that arise due to the effect
of Christoffel symbols. In quaternionic formulation, we also can emphasize
that the space-time curvature path is not only followed by vector
components but also followed by the scalar components and mixed components
of a quaternion variable. If the Christoffel symbol is zero, then
the quaternionic geodesic equations of motion lead to
\begin{align}
\frac{d^{2}P^{\mu}}{ds^{2}}\,\,=\,\, & 0\,,\,\,\,\,\,\,\,\forall\,(\mu=0,1,2,3)\,.\label{eq:72}
\end{align}
This implies that the generalized quaternionic acceleration will be
zero i.e. the uniform velocity of the particle moving in a straight
line (or non-Riemannian space-time).

\subsection{Quaternionic Riemannian Christoffel curvature tensor}

The Riemannian Christoffel curvature tensor is a four-indices tensor,
can be obtained by subtraction of two covariant derivatives of quaternionic
tensors in which indices are interchanged, such that
\begin{align}
T_{\xi\xi;\phi}-T_{\xi\phi;\xi}\,\,=\,\, & \left(R_{\xi\xi\phi}^{\eta}\right)H_{\eta}\,,\label{eq:73}
\end{align}
where
\begin{equation}
R_{\xi\xi\phi}^{\eta}\,\,=\,\,\partial_{\xi}\left(\left\{ \begin{array}{c}
\eta\\
\phi\xi
\end{array}\right\} \right)-\partial_{\phi}\left(\left\{ \begin{array}{c}
\eta\\
\xi\xi
\end{array}\right\} \right)+\left\{ \begin{array}{c}
\varphi\\
\phi\xi
\end{array}\right\} \left\{ \begin{array}{c}
\eta\\
\varphi\xi
\end{array}\right\} -\left\{ \begin{array}{c}
\varphi\\
\xi\xi
\end{array}\right\} \left\{ \begin{array}{c}
\eta\\
\varphi\phi
\end{array}\right\} \label{eq:74}
\end{equation}
is the Riemannian Christoffel curvature tensor for purely quaternionic
scalar field. It is a four-indices tensor in curved space-time which
describes the curvature of manifolds. Further, we can obtain the Riemannian
Christoffel curvature tensor to other quaternionic components given
in equation (\ref{eq:52})-(\ref{eq:58}), i.e.,
\begin{align}
T_{\xi\xi;k}-T_{\xi k;\xi}\,\,=\,\, & \left(R_{\xi\xi k}^{l}\right)H_{l}\,,\label{eq:75}\\
T_{\xi n;\phi}-T_{\xi\phi;n}\,\,=\,\, & \left(R_{\xi n\phi}^{\eta}\right)H_{\eta}\,,\label{eq:76}\\
T_{\xi n;k}-T_{\xi k;n}\,\,=\,\, & \left(R_{\xi nk}^{l}\right)H_{l}\,,\label{eq:77}\\
T_{j\xi;\phi}-T_{j\phi;\xi}\,\,=\,\, & \left(R_{j\xi\phi}^{\eta}\right)H_{\eta}\,,\label{eq:78}\\
T_{j\xi;k}-T_{jk;\xi}\,\,=\,\, & \left(R_{j\xi k}^{l}\right)H_{l}\,,\label{eq:79}\\
T_{jn;\phi}-T_{j\phi;n}\,\,=\,\, & \left(R_{jn\phi}^{\eta}\right)H_{\eta}\,,\label{eq:80}\\
T_{jn;k}-T_{jk;n}\,\,=\,\, & \left(R_{jnk}^{l}\right)H_{l}\,,\label{eq:81}
\end{align}
where the Riemannian Christoffel curvature tensors lead to the following
way as
\begin{align}
R_{\xi\xi k}^{l}\,\,=\,\, & \partial_{\xi}\left(\left\{ \begin{array}{c}
l\\
k\xi
\end{array}\right\} \right)-\partial_{k}\left(\left\{ \begin{array}{c}
l\\
\xi\xi
\end{array}\right\} \right)+\left\{ \begin{array}{c}
i\\
k\xi
\end{array}\right\} \left\{ \begin{array}{c}
l\\
i\xi
\end{array}\right\} -\left\{ \begin{array}{c}
i\\
\xi\xi
\end{array}\right\} \left\{ \begin{array}{c}
l\\
ik
\end{array}\right\} \,,\label{eq:82}\\
R_{\xi n\phi}^{\eta}\,\,=\,\, & \partial_{n}\left(\left\{ \begin{array}{c}
\eta\\
\phi\xi
\end{array}\right\} \right)-\partial_{\phi}\left(\left\{ \begin{array}{c}
\eta\\
\xi n
\end{array}\right\} \right)+\left\{ \begin{array}{c}
\varphi\\
\phi\xi
\end{array}\right\} \left\{ \begin{array}{c}
\eta\\
\varphi n
\end{array}\right\} -\left\{ \begin{array}{c}
\varphi\\
n\xi
\end{array}\right\} \left\{ \begin{array}{c}
\eta\\
\varphi\phi
\end{array}\right\} \,,\label{eq:83}\\
R_{\xi nk}^{l}\,\,=\,\, & \partial_{n}\left(\left\{ \begin{array}{c}
l\\
k\xi
\end{array}\right\} \right)-\partial_{k}\left(\left\{ \begin{array}{c}
l\\
\xi n
\end{array}\right\} \right)+\left\{ \begin{array}{c}
i\\
k\xi
\end{array}\right\} \left\{ \begin{array}{c}
l\\
in
\end{array}\right\} -\left\{ \begin{array}{c}
i\\
n\xi
\end{array}\right\} \left\{ \begin{array}{c}
l\\
ik
\end{array}\right\} \,,\label{eq:84}\\
R_{j\xi\phi}^{\eta}\,\,=\,\, & \partial_{\xi}\left(\left\{ \begin{array}{c}
\eta\\
\phi j
\end{array}\right\} \right)-\partial_{\phi}\left(\left\{ \begin{array}{c}
\eta\\
j\xi
\end{array}\right\} \right)+\left\{ \begin{array}{c}
\varphi\\
\phi j
\end{array}\right\} \left\{ \begin{array}{c}
\eta\\
\varphi\xi
\end{array}\right\} -\left\{ \begin{array}{c}
\varphi\\
\xi j
\end{array}\right\} \left\{ \begin{array}{c}
\eta\\
\varphi\phi
\end{array}\right\} \,,\label{eq:85}\\
R_{j\xi k}^{l}\,\,=\,\, & \partial_{\xi}\left(\left\{ \begin{array}{c}
l\\
kj
\end{array}\right\} \right)-\partial_{k}\left(\left\{ \begin{array}{c}
l\\
j\xi
\end{array}\right\} \right)+\left\{ \begin{array}{c}
i\\
kj
\end{array}\right\} \left\{ \begin{array}{c}
l\\
i\xi
\end{array}\right\} -\left\{ \begin{array}{c}
i\\
\xi j
\end{array}\right\} \left\{ \begin{array}{c}
l\\
ik
\end{array}\right\} \,,\label{eq:86}\\
R_{jn\phi}^{\eta}\,\,=\,\, & \partial_{n}\left(\left\{ \begin{array}{c}
\eta\\
\phi j
\end{array}\right\} \right)-\partial_{\phi}\left(\left\{ \begin{array}{c}
\eta\\
jn
\end{array}\right\} \right)+\left\{ \begin{array}{c}
\varphi\\
\phi j
\end{array}\right\} \left\{ \begin{array}{c}
\eta\\
\varphi n
\end{array}\right\} -\left\{ \begin{array}{c}
\varphi\\
nj
\end{array}\right\} \left\{ \begin{array}{c}
\eta\\
\varphi\phi
\end{array}\right\} \,,\label{eq:87}\\
R_{jnk}^{l}\,\,=\,\, & \partial_{n}\left(\left\{ \begin{array}{c}
l\\
kj
\end{array}\right\} \right)-\partial_{k}\left(\left\{ \begin{array}{c}
l\\
jn
\end{array}\right\} \right)+\left\{ \begin{array}{c}
i\\
kj
\end{array}\right\} \left\{ \begin{array}{c}
l\\
in
\end{array}\right\} -\left\{ \begin{array}{c}
i\\
nj
\end{array}\right\} \left\{ \begin{array}{c}
l\\
ik
\end{array}\right\} \,.\label{eq:88}
\end{align}
Here, the quaternionic Riemannian tensor keeps track that how much
scalar and vector components of quaternion change when we propagate
parallel along with a small parallelogram. If the value of quaternionic
Riemannian Christoffel curvature is zero then the quaternionic curved
space-time is converted into flat space-time. On the other hand, the
quaternionic Ricci tensor is an important contraction of quaternionic
Riemannian Christoffel tensor which explains the volume changes when
an object parallels transport along a geodesic. In this case
\begin{align}
g_{h\eta}R_{\xi n\phi}^{\eta}\,\,=\,\, & R_{h\xi n\phi}\nonumber \\
=\,\, & \partial_{n}\left(g_{h\eta}\left\{ \begin{array}{c}
\eta\\
\phi\xi
\end{array}\right\} \right)-\partial_{\phi}\left(g_{h\eta}\left\{ \begin{array}{c}
\eta\\
\xi n
\end{array}\right\} \right)+g_{h\eta}\left\{ \begin{array}{c}
\varphi\\
\phi\xi
\end{array}\right\} \left\{ \begin{array}{c}
\eta\\
\varphi n
\end{array}\right\} \nonumber \\
\,\, & -g_{h\eta}\left\{ \begin{array}{c}
\varphi\\
n\xi
\end{array}\right\} \left\{ \begin{array}{c}
\eta\\
\varphi\phi
\end{array}\right\} \,,\label{eq:89}
\end{align}
which can be simplified in terms of a metric tensor as
\begin{align}
R_{h\xi n\phi}\,\,=\,\, & \frac{1}{2}\left[\frac{\partial^{2}g_{h\phi}}{\partial P^{n}\partial P^{\xi}}+\frac{\partial^{2}g_{\xi n}}{\partial P^{h}\partial P^{\phi}}-\frac{\partial^{2}g_{\xi\phi}}{\partial P^{n}\partial P^{h}}-\frac{\partial^{2}g_{nh}}{\partial P^{\phi}\partial P^{\xi}}\right]\nonumber \\
+\,\, & g_{\eta h}\left\{ \begin{array}{c}
\varphi\\
\phi\xi
\end{array}\right\} \left\{ \begin{array}{c}
\eta\\
\phi n
\end{array}\right\} -g_{\eta h}\left\{ \begin{array}{c}
\varphi\\
n\xi
\end{array}\right\} \left\{ \begin{array}{c}
\eta\\
\varphi\phi
\end{array}\right\} \,.\label{eq:90}
\end{align}
Now, contracting equation (\ref{eq:90}) with the metric tensor as
$g^{hn}R_{h\xi n\phi}=\,\,R_{\xi\phi}$, where $R_{\xi\phi}$ is a
quaternionic form of Ricci tensor in scalar field, so that
\begin{align}
R_{\xi\phi}\,\,=\,\, & \partial_{n}\left(\left\{ \begin{array}{c}
n\\
\phi\xi
\end{array}\right\} \right)-\partial_{\phi}\left(\left\{ \begin{array}{c}
n\\
\xi n
\end{array}\right\} \right)+\left\{ \begin{array}{c}
\varphi\\
\phi\xi
\end{array}\right\} \left\{ \begin{array}{c}
n\\
\varphi n
\end{array}\right\} -\left\{ \begin{array}{c}
\varphi\\
n\xi
\end{array}\right\} \left\{ \begin{array}{c}
n\\
\varphi\phi
\end{array}\right\} \,.\label{eq:91}
\end{align}
Similarly, to involving quaternionic vector field in addition to scalar
field, also in pure vector field the Ricci tensors become
\begin{align}
R_{\xi k}\,\,=\,\, & \partial_{n}\left(\left\{ \begin{array}{c}
n\\
k\xi
\end{array}\right\} \right)-\partial_{k}\left(\left\{ \begin{array}{c}
n\\
\xi n
\end{array}\right\} \right)+\left\{ \begin{array}{c}
i\\
k\xi
\end{array}\right\} \left\{ \begin{array}{c}
n\\
in
\end{array}\right\} -\left\{ \begin{array}{c}
i\\
n\xi
\end{array}\right\} \left\{ \begin{array}{c}
n\\
ik
\end{array}\right\} \,,\label{eq:92}\\
R_{jk}\,\,=\,\, & \partial_{n}\left(\left\{ \begin{array}{c}
n\\
kj
\end{array}\right\} \right)-\partial_{k}\left(\left\{ \begin{array}{c}
n\\
jn
\end{array}\right\} \right)+\left\{ \begin{array}{c}
i\\
kj
\end{array}\right\} \left\{ \begin{array}{c}
n\\
in
\end{array}\right\} -\left\{ \begin{array}{c}
i\\
nj
\end{array}\right\} \left\{ \begin{array}{c}
n\\
ik
\end{array}\right\} \,.\label{eq:93}
\end{align}
We should notice that the Riemannian tensor ($R_{h\xi n\phi}$) is
symmetric to first with third indices and second with fourth indices
\cite{key-3}. Thus the Ricci tensor is also symmetric. The quaternionic
form of Ricci tensor can further contract with metric tensor and obtain
scalar curvature tensor ($R$) of rank zero, such that $g^{\xi\phi}R_{\xi\phi}=\,\,R\,.$

\subsection{Quaternionic Einstein-field like equation}

To express the quaternionic Einstein field like equation, let us start
with the Poisson equation as
\begin{align}
\nabla^{2}\Phi\,\,=\,\, & 4\pi G\rho\,,\label{eq:94}
\end{align}
where $\Phi$ is the Newtonian gravitational potential, $G$ is gravitational
constant and $\rho$ is the mass density. Moreover, in metric tensor
form \cite{key-45}, we have $\nabla^{2}\Phi=-\frac{1}{2}\nabla^{2}g_{00}$.
Thus
\begin{align}
\nabla^{2}g_{00}\,\,=\,\, & -8\pi G\rho\,.\label{eq:95}
\end{align}
It should be noticed that in equation (\ref{eq:95}) we introduced
the quaternionic form of scalar metric tensor ($g_{\xi\xi}$) in place
of Newtonian potential ($\Phi$) and quaternionic mass density ($T_{\xi\xi}$)
in place of ($\rho$). Generally, the quaternionic energy-momentum
tensor can be written as
\begin{align}
T^{\mu\nu}\,\,\simeq\,\, & \left[\begin{array}{cccc}
T^{\xi\xi} & T^{\xi1} & T^{\xi2} & T^{\xi3}\\
T^{1\xi} & T^{11} & T^{12} & T^{13}\\
T^{2\xi} & T^{21} & T^{22} & T^{23}\\
T^{3\xi} & T^{31} & T^{32} & T^{33}
\end{array}\right]\,.\label{eq:96}
\end{align}
Therefore, equation (\ref{eq:95}) leads to $\nabla^{2}g_{\xi\xi}\,=\,-8\pi GT_{\xi\xi}.$
Since the quaternionic Ricci tensor can also be written in the form
of a double derivative of the quaternionic metric tensor. So, we get
\begin{align}
R_{\xi\xi}\,\,=\,\, & -8\pi GT_{\xi\xi}\,.\label{eq:97}
\end{align}
Further, the rest components of quaternionic Riemannian Christoffel
curvature tensors can express as
\begin{align}
R_{\xi1}\,\,=\, & -8\pi GT_{\xi1}\,,\,\,\,\,\,\,\,\,R_{\xi2}\,\,=\,-8\pi GT_{\xi2}\,,\,\,\,\,\,\,\,\,R_{\xi3}\,\,=\,-8\pi GT_{\xi3}\,,\nonumber \\
R_{11}\,\,=\, & -8\pi GT_{11}\,,\,\,\,\,\,\,\,R_{22}\,\,=\,-8\pi GT_{22}\,,\,\,\,\,\,\,\,R_{33}\,\,=\,-8\pi GT_{33}\,,\nonumber \\
R_{12}\,\,=\, & -8\pi GT_{12}\,,\,\,\,\,\,\,\,R_{23}\,\,=\,-8\pi GT_{23}\,,\,\,\,\,\,\,\,R_{31}\,\,=\,-8\pi GT_{31}\,,\label{eq:98}
\end{align}
where the other components i.e. $R_{1\xi},\,R_{2\xi},\,R_{3\xi},\,R_{21},\,R_{32},\,R_{13}$
are symmetrical to $R_{\xi1},\,R_{2\xi},\,R_{3\xi},$$\,R_{12},\,R_{23},\,R_{31}$,
respectively. Therefore, we may write a generalized quaternionic form
of Einstein field-like equation as $R_{\mu\nu}\,\,=\,\,-8\pi GT_{\mu\nu}\,.$
Since the quaternionic energy-momentum tensor is conserved quantity
but quaternionic Ricci tensor is not, so it should be divergence-free.
Thus, by using Bianchi identity $R_{\alpha\mu\beta\gamma;\,\sigma}+R_{\alpha\mu\gamma\sigma;\,\beta}+R_{\alpha\mu\sigma\beta;\,\gamma}\,\,=\,\,0$,
the quaternionic Einstein field-like equation can be written as $R_{\mu\nu}-\frac{1}{2}\,g_{\mu\nu}R\,\,=\,\,-8\pi GT_{\mu\nu}$.
Here, we can see that the right-hand part shows quaternionic energy
and matter, while the left-hand part shows the quaternionic geometry.
Thus the quaternionic formalism is perfectly fitted to describe a
new approach in the dynamics of a particle in curved space-time.

\section{Conclusion}

In Riemannian geometry, the curvature of space-time is directly connected
to the energy and momentum of whatever matter and radiation are present.
Generally, the space-time can be shown using simple experiments following
the free-fall trajectories of different test particles, the result
of transporting space-time vectors that can denote a particle's velocity
(time-like vectors) will vary with the particle's trajectory. The
origin of the idea of this space-time curvature is a mathematical
concept. Thus the four-dimensional curvature space-time has been visualized
by a quaternionic algebra. The transformation rules for two quaternionic
frames have been discussed. Interestingly, we discussed that the effect
of curvature space-time arises due to the quaternionic Christoffel
symbol. Also, it is concluded that the curved path of freely moving
particles is effectively explained by the quaternionic geodesic equation.
The Christoffel symbol has been established in terms of the derivative
of quaternionic metric tensor while the gravitational potential in
tensorial form is represented by the metric tensor. It has been confirmed
that the volume element of an object obeying the geodesic path changes
due to the contraction of the quaternionic Riemannian tensor while
the contraction of the quaternionic Ricci tensor gives a scalar quantity
representing the magnitude of change in volume of the object. We also
have been discussed the quaternionic Einstein field-like equation
for gravitation that relates the geometry of quaternionic space-time
with the distribution of matter within it. The simplification of quaternionic
formalism will be achieved by the approximation as quaternionic flat
space-time with a small deviation, which leads to the linearized Einstein
field equation. These quaternionic equations may be used to study
the phenomena such as gravitational waves.

\end{document}